


\font\bigbold=cmbx12
\font\ninerm=cmr9
\font\eightrm=cmr8
\font\sixrm=cmr6
\font\fiverm=cmr5
\font\ninebf=cmbx9
\font\eightbf=cmbx8
\font\sixbf=cmbx6
\font\fivebf=cmbx5
\font\ninei=cmmi9  \skewchar\ninei='177
\font\eighti=cmmi8  \skewchar\eighti='177
\font\sixi=cmmi6    \skewchar\sixi='177
\font\fivei=cmmi5
\font\ninesy=cmsy9 \skewchar\ninesy='60
\font\eightsy=cmsy8 \skewchar\eightsy='60
\font\sixsy=cmsy6   \skewchar\sixsy='60
\font\fivesy=cmsy5
\font\nineit=cmti9
\font\eightit=cmti8
\font\ninesl=cmsl9
\font\eightsl=cmsl8
\font\ninett=cmtt9
\font\eighttt=cmtt8
\font\tenfrak=eufm10
\font\ninefrak=eufm9
\font\eightfrak=eufm8
\font\sevenfrak=eufm7
\font\fivefrak=eufm5
\font\tenbb=msbm10
\font\ninebb=msbm9
\font\eightbb=msbm8
\font\sevenbb=msbm7
\font\fivebb=msbm5
\font\tensmc=cmcsc10


\newfam\bbfam
\textfont\bbfam=\tenbb
\scriptfont\bbfam=\sevenbb
\scriptscriptfont\bbfam=\fivebb
\def\Bbb{\fam\bbfam}

\newfam\frakfam
\textfont\frakfam=\tenfrak
\scriptfont\frakfam=\sevenfrak
\scriptscriptfont\frakfam=\fivefrak
\def\frak{\fam\frakfam}

\def\smc{\tensmc}


\def\eightpoint{%
\textfont0=\eightrm   \scriptfont0=\sixrm
\scriptscriptfont0=\fiverm  \def\rm{\fam0\eightrm}%
\textfont1=\eighti   \scriptfont1=\sixi
\scriptscriptfont1=\fivei  \def\oldstyle{\fam1\eighti}%
\textfont2=\eightsy   \scriptfont2=\sixsy
\scriptscriptfont2=\fivesy
\textfont\itfam=\eightit  \def\it{\fam\itfam\eightit}%
\textfont\slfam=\eightsl  \def\sl{\fam\slfam\eightsl}%
\textfont\ttfam=\eighttt  \def\tt{\fam\ttfam\eighttt}%
\textfont\frakfam=\eightfrak \def\frak{\fam\frakfam\eightfrak}%
\textfont\bbfam=\eightbb  \def\Bbb{\fam\bbfam\eightbb}%
\textfont\bffam=\eightbf   \scriptfont\bffam=\sixbf
\scriptscriptfont\bffam=\fivebf  \def\bf{\fam\bffam\eightbf}%
\abovedisplayskip=9pt plus 2pt minus 6pt
\belowdisplayskip=\abovedisplayskip
\abovedisplayshortskip=0pt plus 2pt
\belowdisplayshortskip=5pt plus2pt minus 3pt
\smallskipamount=2pt plus 1pt minus 1pt
\medskipamount=4pt plus 2pt minus 2pt
\bigskipamount=9pt plus4pt minus 4pt
\setbox\strutbox=\hbox{\vrule height 7pt depth 2pt width 0pt}%
\normalbaselineskip=9pt \normalbaselines
\rm}


\def\ninepoint{%
\textfont0=\ninerm   \scriptfont0=\sixrm
\scriptscriptfont0=\fiverm  \def\rm{\fam0\ninerm}%
\textfont1=\ninei   \scriptfont1=\sixi
\scriptscriptfont1=\fivei  \def\oldstyle{\fam1\ninei}%
\textfont2=\ninesy   \scriptfont2=\sixsy
\scriptscriptfont2=\fivesy
\textfont\itfam=\nineit  \def\it{\fam\itfam\nineit}%
\textfont\slfam=\ninesl  \def\sl{\fam\slfam\ninesl}%
\textfont\ttfam=\ninett  \def\tt{\fam\ttfam\ninett}%
\textfont\frakfam=\ninefrak \def\frak{\fam\frakfam\ninefrak}%
\textfont\bbfam=\ninebb  \def\Bbb{\fam\bbfam\ninebb}%
\textfont\bffam=\ninebf   \scriptfont\bffam=\sixbf
\scriptscriptfont\bffam=\fivebf  \def\bf{\fam\bffam\ninebf}%
\abovedisplayskip=10pt plus 2pt minus 6pt
\belowdisplayskip=\abovedisplayskip
\abovedisplayshortskip=0pt plus 2pt
\belowdisplayshortskip=5pt plus2pt minus 3pt
\smallskipamount=2pt plus 1pt minus 1pt
\medskipamount=4pt plus 2pt minus 2pt
\bigskipamount=10pt plus4pt minus 4pt
\setbox\strutbox=\hbox{\vrule height 7pt depth 2pt width 0pt}%
\normalbaselineskip=10pt \normalbaselines
\rm}


\def\pagewidth#1{\hsize= #1}
\def\pageheight#1{\vsize= #1}
\def\hcorrection#1{\advance\hoffset by #1}
\def\vcorrection#1{\advance\voffset by #1}

\newif\iftitlepage   \titlepagetrue               
\newtoks\titlepagefoot     \titlepagefoot={\hfil} 
\newtoks\otherpagesfoot    \otherpagesfoot={\hfil\tenrm\folio\hfil}
\footline={\iftitlepage\the\titlepagefoot\global\titlepagefalse
           \else\the\otherpagesfoot\fi}

\font\extra=cmss10 scaled \magstep0
\setbox1 = \hbox{{{\extra R}}}
\setbox2 = \hbox{{{\extra I}}}
\setbox3 = \hbox{{{\extra C}}}
\setbox4 = \hbox{{{\extra Z}}}
\setbox5 = \hbox{{{\extra N}}}

\def\RRR{{{\extra R}}\hskip-\wd1\hskip2.0
   true pt{{\extra I}}\hskip-\wd2
\hskip-2.0 true pt\hskip\wd1}
\def\Real{\hbox{{\extra\RRR}}}    

\def\CCC{{{\extra C}}\hskip-\wd3\hskip 2.5 true pt{{\extra I}}
\hskip-\wd2\hskip-2.5 true pt\hskip\wd3}
\def\Complex{\hbox{{\extra\CCC}}\!\!}   

\def\ZZZ{{{\extra Z}}\hskip-\wd4\hskip 2.5 true pt{{\extra Z}}}
\def\Zed{\hbox{{\extra\ZZZ}}}       




\def\Z{{\Zed}}
\def\R{{\Real}}
\def\C{{\Complex}}

\def\frac#1#2{{#1\over#2}}

\def\({\left(}
\def\){\right)}

\def\pmb#1{\setbox0=\hbox{$#1$}%
   \kern-.025em\copy0\kern-\wd0
   \kern.05em\copy0\kern-\wd0
   \kern-.025em\raise.0433em\box0 }


\def\abstract#1{{\parindent=30pt\narrower\noindent\ninepoint\openup
2pt #1\par}}


\newcount\notenumber\notenumber=1
\def\note#1
{\unskip\footnote{$^{\the\notenumber}$}
{\eightpoint\openup 1pt #1}
\global\advance\notenumber by 1}


\global\newcount\secno \global\secno=0
\global\newcount\meqno \global\meqno=1
\global\newcount\appno \global\appno=0
\newwrite\eqmac
\def\romappno{\ifcase\appno\or A\or B\or C\or D\or E\or F\or G\or H
\or I\or J\or K\or L\or M\or N\or O\or P\or Q\or R\or S\or T\or U\or
V\or W\or X\or Y\or Z\fi}
\def\eqn#1{
        \ifnum\secno>0
            \eqno(\the\secno.\the\meqno)\xdef#1{\the\secno.\the\meqno}
          \else\ifnum\appno>0
            \eqno({\rm\romappno}.\the\meqno)\xdef#1{{\rm\romappno}.\the=
\meqno}
          \else
            \eqno(\the\meqno)\xdef#1{\the\meqno}
          \fi
        \fi
\global\advance\meqno by1 }


\global\newcount\refno
\global\refno=1 \newwrite\reffile
\newwrite\refmac
\newlinechar=`\^^J
\def\ref#1#2{\the\refno\nref#1{#2}}
\def\nref#1#2{\xdef#1{\the\refno}
\ifnum\refno=1\immediate\openout\reffile=refs.tmp\fi
\immediate\write\reffile{
     \noexpand\item{[\noexpand#1]\ }#2\noexpand\nobreak.}
     \immediate\write\refmac{\def\noexpand#1{\the\refno}}
   \global\advance\refno by1}
\def\semi{;\hfil\noexpand\break ^^J}
\def\nl{\hfil\noexpand\break ^^J}
\def\refn#1#2{\nref#1{#2}}

\def\jpA#1#2#3{{\it J.  Phys.} {\bf A{#1}} (19{#2}) #3}
\def\ijmp#1#2#3{{\it Int.  J.  Mod.  Phys.} {\bf A{#1}} (19{#2}) #3}

\def\plA#1#2#3{{\it Phys.  Lett.} {\bf {#1}A} (19{#2}) #3}

\def\prD#1#2#3{{\it Phys.  Rev.} {\bf D{#1}} (19{#2}) #3}
\def\prl#1#2#3{{\it Phys.  Rev.  Lett.} {\bf #1} (19{#2}) #3}


{

\refn\AGHH
{S. Albeverio, F. Gesztesy, R. H{\o}egh-Krohn and H. Holden,
\lq\lq Solvable Models in Quantum Mechanics\rq\rq,
Springer, New York, 1988}

\refn\Seba
{P. \v{S}eba,
{\it Czech. J. Phys.} {\bf 36} (1986) 667}

\refn\CH
{P.R. Chernoff and R.J. Hughes,
{\it J. Funct. Anal.} {\bf 111} (1993) 97}

\refn\ABD
{S. Albeverio, Z. Brze\'{z}niak and L. Dabrowski,
{\it J. Funct. Anal.} {\bf 130} (1995) 220}

\refn\Car
{M. Carreau,
\jpA{26}{93}{427}}

\refn\RT
{J.M. Rom\'{a}n and R. Tarrach,
\jpA{29}{96}{6073}}

\refn\CS
{T. Cheon and T. Shigehara,
\plA{243}{98}{111}}

\refn\AN
{S. Albeverio and L. Nizhnik,
{\sl Approximation of general Zero-Range Potentials},
Uni. Bonn Preprint no.585 (1999)}

\refn\Schulman
{L.S. Schulman, 
\lq\lq Techniques and Applications of
Path Integration\rq\rq, John Wiley \& Sons, New York, 1981}

\refn\Schiff
{L. Schiff,
\lq\lq Quantum Mechanics\rq\rq, 3rd ed.,
McGraw-Hill, Tokyo, 1968}

\refn\FT
{T. F\"{u}l\"{o}p and I. Tsutsui,
in preparation}

\refn\RS
{M. Reed and B. Simon, 
\lq\lq Methods of Modern Mathematical Physics\rq\rq, Vol.I, II,
Academic Press, New York, 1980}

\refn\Kleinert
{H. Kleinert, 
\lq\lq Path Integrals in Quantum Mechanics, Statistics and 
Polymer Physics\rq\rq, 2nd ed., 
World Scientific, Singapore, 1995}

\refn\CMS
{T.E. Clark, R. Menikoff and D.H. Sharp,
\prD{22}{80}{3012}}

\refn\FG
{E. Farhi and S. Gutmann,
\ijmp{5}{90}{3029}}

\refn\ADK
{S. Albeverio, L. Dabrowski and P. Kurasov,
{\it Lett. Math. Phys.} {\bf 45} (1998) 33}

\refn\CFG
{M. Carreau, E. Farhi and S. Gutmann,
\prD{42}{90}{1194}}

\refn\Buslaev
{V.S. Buslaev,
in \lq\lq Topics in Mathematical Physics, Vol.2, 
Spectral theory and Diffraction\rq\rq,
Ed. M. Sh. Birman, Consultants Bureau, New York, 1967}

\refn\KM
{J.B. Keller and D.W. McLaughlin,
{\it Am. Math. Monthly} {\bf 82} (1975) 451}

\refn\Jackiw
{R. Jackiw, 
in \lq\lq Current Algebras and Anomalies\rq\rq,
World Scientific, Singapore, 1985}

\refn\Isham
{C.J. Isham, 
in \lq\lq Relativity, Groups and Topology
II\rq\rq, 
Eds. B.S. DeWitt and R. Stora, 
\nl North-Holland,
Amsterdam, 1984}

\refn\Cheon
{T. Cheon,
\plA{248}{98}{285}}

\refn\CSb
{T. Cheon and T. Shigehara,
\prl{82}{99}{2536}}

\refn\Jackiwb
{R. Jackiw, 
in \lq\lq Diverse topics in 
Theoretical and Mathematical Physics\rq\rq,
World Scientific, Singapore, 1995}

}


\def\ve{\vfill\eject}




\pageheight{23cm}
\pagewidth{14.8cm}
\hcorrection{0mm}
\magnification= \magstep1
\def\bsk{%
\baselineskip= 16.8pt plus 1pt minus 1pt}
\parskip=5pt plus 1pt minus 1pt
\tolerance 6000


\null

{
\leftskip=100mm
\hfill\break
KEK Preprint 99-115
\hfill\break
\par}

\smallskip
\vfill
{\baselineskip=18pt

\centerline{\bigbold A Free Particle on a Circle
with Point Interaction}

\vskip 30pt

\centerline{
\smc 
Tam\'{a}s F\"{u}l\"{o}p\note
{E-mail:\quad fulopt@poe.elte.hu}
}

\vskip 5pt
{
\baselineskip=13pt
\centerline{\it Institute for Theoretical Physics}
\centerline{\it Roland E\"{o}tv\"{o}s University}
\centerline{\it H-1117 Budapest, P\'{a}zm\'{a}ny 
P. s\'{e}t\'{a}ny 1/A, Hungary}
}

\vskip 5pt
\centerline{\rm and}
\vskip 5pt

\centerline{
\smc
Izumi Tsutsui\note
{E-mail:\quad izumi.tsutsui@kek.jp}
}

\vskip 5pt

{
\baselineskip=13pt
\centerline{\it
Institute of Particle and Nuclear Studies}
\centerline{\it
High Energy Accelerator Research Organization (KEK),
Tanashi Branch}
\centerline{\it Tokyo 188-8501, Japan}
}

\vskip 70pt

\abstract{%
{\bf Abstract.}\quad
The quantum dynamics of a free particle on a circle
with point interaction is described by a $U(2)$ family 
of self-adjoint Hamiltonians.  
We provide a classification of the family by introducing 
a number of subfamilies and thereby analyze
the spectral structure in detail.  We find that
the spectrum depends on a subset of $U(2)$ parameters 
rather than the entire $U(2)$ needed for the Hamiltonians, 
and that in particular there exists a subfamily in $U(2)$
where the spectrum becomes parameter-independent.
We also show that, in some specific cases, 
the WKB semiclassical
approximation becomes exact (modulo phases) for the system.
}

\bigskip
{\ninepoint
PACS codes: 03.65.Db; 03.65.Ge; 03.65.Sq \hfill\break
}


\pageheight{23cm}
\pagewidth{15.7cm}
\hcorrection{-1mm}
\magnification= \magstep1
\def\bsk{%
\baselineskip= 14.5pt plus 1pt minus 1pt}
\parskip=5pt plus 1pt minus 1pt
\tolerance 8000
\bsk


\ve
\secno=1 \meqno=1


\bigskip
\noindent{\bf 1. Introduction}
\medskip

Systems with point interaction form an important
class of solvable models in quantum mechanics, allowing for
a variety of applications in physics (see [\AGHH] and references
therein).  These systems
are governed by Hamiltonian operators
given by a {\it point perturbation} of the Laplacian on $\R^n$,
that is, the self-adjoint Laplacian operators 
on $\R^n$ with one point removed.
In one dimension, it has been recognized [\Seba, \AGHH] 
that the perturbation
yields a $U(2)$ family of self-adjoint operators 
on $\R^1\!\setminus\!\{0\}$, which implies that there is
a $U(2)$ family of
distinct point interactions possible there.
The physical properties of the family has 
been investigated in [\CH], and more fully in [\ABD]
including the spectra, time-dependent fundamental solutions
and scattering matrices.
Recently, rapid
technological advances in microscopic devices spurred a new 
interest in realizing these point interactions 
using regulated potentials [\Car, \RT, \CS, \AN].

The aim of the present paper is twofold.  First, we 
provide a study on point interaction on a circle 
$S^1$ analogous to that made on a line $\R^1$. 
This is interesting for the reason
that the difference in 
self-adjoint operators on $S^1\!\setminus\!\{0\}$
--- which are again characterized by $U(2)$ --- 
arises in discrete energy spectrum rather than 
in the scattering amplitude as 
on $\R^1\!\setminus\!\{0\}$ 
whose positive energy spectrum is always continuous.  
We shall see that the spectra of the $U(2)$ family exhibit 
remarkable characteristics; in particular, the spectral 
space forms a subspace of $U(2)$, rather than the whole 
$U(2)$ as apparently suggested by the theory of 
self-adjoint extensions.
In a subfamily of $U(2)$ (called `separated subfamily'), 
the circle becomes
equivalent to a box with nontrivial boundary conditions,
and hence our result is also relevant to a box 
if such boundary conditions are available.  
Secondly, we show that in some specific cases
our system on $S^1\!\setminus\!\{0\}$ 
admits the interpretation that
the WKB semicalssical
approximation is exact up to a phase which is 
determined by the point interaction.
In fact, the WKB exactness of a box system 
has been pointed out earlier
in the path-integral for perfectly 
reflecting walls [\Schulman], and in this respect
our cases provide a generalization of the particular
case previously considered. 

This paper is organized as follows.  In sect.~2 we review
briefly the system of a particle confined 
to a half line 
where the basic properties mentioned above, such as 
the dependence of the spectrum
on the allowed self-adjoint extensions and the WKB exactness
in the transition amplitude, can easily be seen.   
We then present our result on $S^1\!\setminus\!\{0\}$ 
in sect.~3 by considering 
a particle confined to a box under the most general
self-adjoint extensions of the Laplacian, a setup which is  
equivalent to a circle with point interaction.  For our 
convenience we introduce a number of subfamilies defined 
within the $U(2)$ family.  These subfamilies 
are characterized by distinguished physical properties, wherein
the spectra and the WKB exactness are discussed separately.

\secno=2 \meqno=1


\bigskip
\noindent{\bf 2. Particle on a Half Line}
\medskip

We begin by discussing a particle restricted to move
on the half line $x \ge 0$.  The system will be given by   
placing an `infinite'
potential wall at $x = 0$ on a line, and
in quantum mechanics one conventionally imposes the 
vanishing boundary condition $\psi(0) = 0$ for 
wavefunctions.  This is
based on the observation (see, {\it e.g.}, [\Schiff]) 
that the boundary condition
that arises under a finite constant wall $V_0$  
at $x = 0$ 
reduces to the conventional one
in the limit $V_0 \rightarrow \infty$.    
This, however, is too restrictive to specify the boundary
condition, because other limits that also confine
the particle in the half line can lead to
boundary conditions which are different from 
the conventional one [\FT].
It is therefore safe to say that the only requirement for
the particle to be confined on the half line
is that the probability current $j(x)$, not the
wave function, vanish at the wall,
$$
j(0) = 0, \qquad \hbox{where} \quad
j(x) = - {{i\hbar}\over{2m}}\left(
(\psi^*)'\psi - \psi^* \psi' \right)(x)\ .
\eqn\ctwall
$$
The condition (\ctwall) is then satisfied if
$$
\psi(0) + L \, \psi'(0) = 0\ ,
\eqn\bcwall
$$
with $L \in \R$ being an arbitrary constant
of length dimension, or $\psi'(0) = 0$
which may be thought of the case $L = \infty$ in (\bcwall).
The combined space of the solutions 
$\R \cup \{\infty\} \simeq U(1)$ to the condition (\ctwall)
may thus be parametrized by $L := L_0 \cot\phi$ with 
an angle $\phi \in [0, \pi)$ and a nonvanishing 
constant $L_0$.
The same $U(1)$ family can be obtained from
the theory of self-adjoint extensions
applied to the free Hamiltonian 
$H = - {{\hbar^2}\over{2m}}{{d^2}\over{dx^2}}$ 
restricted to the half line [\RS], 
because self-adjointness
implies probability conservation in the system 
as a whole.\note{
Here one adopts the local
conservation $j(\infty) = 0$ in addition to (\ctwall)
on the ground that the more general global conservation, 
$j(0) = j(\infty)$, will never be realized
physically in practice unless $j(\infty) = 0$.
In parallel, the possibility for a nonzero $j(\infty)$ is excluded by the
results of self-adjoint extension theory, too.}

We then observe that, in addition to
the continuum spectrum for positive energy $E_k = \hbar^2 k^2/2m$
with $k > 0$ formed by the states, 
$$
\psi_k^L(x) = {1\over{\sqrt{2\pi}}}
\left[
e^{-ikx} + \left({{1-ikL}\over{1+ikL}}\right) e^{ikx}
\right]\ ,
\eqn\iowall
$$ 
the system admits
a normalizable bound state 
$$
\psi_{\rm bound}^L(x) = \sqrt{{2\over{L}}} \,
e^{-{x\over{L}}}\ ,
\eqn\bswall
$$
for $L > 0$ with energy 
$E_{\rm B} = - \hbar^2/2m L^2$.
The bound state (\bswall) disappears either at 
$L = 0$ or $L = \infty$ where the boundary condition becomes
$\psi(0) = 0$ or $\psi'(0) = 0$.

The two exceptional 
cases, $L = 0$ and $L = \infty$, which has no 
scale parameter are of particular interest,
because one can then evaluate
the transition amplitude (Feynman kernel)
in closed form.  Indeed, from (\iowall)
the Feynman kernel
$K(b, T; a, 0)$, 
which describes the transition from 
$x = a$ at $t = 0$ to $x = b$ at $t = T$, 
is found to be
$$
\eqalign{
K(b, T; a, 0) 
&= \int^\infty_0 dk\,  e^{-{i\over\hbar}E_k T}
\psi_k^L(b)\,(\psi_k^L)^*(a) \cr
&= \sqrt{m\over{2\pi i \hbar T}}
\left(e^{{i\over\hbar} {m\over{2T}}(b - a)^2}
   \mp  e^{{i\over\hbar} {m\over{2T}}(b + a)^2}\right)\ ,
}
\eqn\fkwall
$$
where the `$-$'-sign in (\fkwall) is for
$L = 0$ whereas the `$+$'-sign is for $L = \infty$.
In the path-integral point of view, 
the result (\fkwall) may be interepreted as being
a sum of contributions from the two possible
classical
trajectories, the direct path from $a$ to $b$ and
the reflected path which starts from $a$ and 
hits the wall before reaching $b$.
We therefore see that, modulo 
the phase shift by $\pi$ or $0$ acquired
at the reflection, 
the WKB semiclassical approximation becomes exact in
the two cases of the present system.  We remark
that the phase shift may also be explained by the
corresponding shift at the reflection
in the classical action, 
if one considers a proper sequence of regulated potentials
that leads to the infinite wall in the limit. 
(The path-integral
treatise on the half line has been discussed earlier in
[\Schulman, \Kleinert] for $L = 0$ and in [\CMS, \FG] for 
generic $L$.)

\secno=3 \meqno=1


\bigskip
\noindent{\bf 3. Particle in a Box
or on a Circle with Point Interaction}
\medskip

We now consider the system of an infinite potential
well, {\it i.e.}, a free particle confined in a box of infinite
potential walls at $x = 0$ and $x = l$.   
It is known that there exists a $U(2)$ family of 
self-adjoint Hamiltonians for the system.  
Using a matrix $U \in U(2)$ and
the identity matrix $I$, the different self-adjoint
extensions in the family can be described by
the boundary conditions, 
$$
(U - I)\Psi + iL_0\, (U + I) \Psi' = 0,
\qquad 
\Psi := \pmatrix{\psi(0) \cr \psi(l) \cr}, \quad
\Psi' := \pmatrix{\psi'(0) \cr - \psi'(l) \cr}\ , 
\eqn\safam
$$
where $L_0$ is again a constant of length dimension.
As before, one can derive these boundary conditions (\safam)
by requiring probability conservation of the entire system.  
To see this, note first that with $\Psi^\dagger = (\Psi^*)^t$ 
the probability conservation, 
$$
j(0) - j(l) = 0\ ,
\eqn\globalcons
$$
is equivalent to
$\Psi^\dagger \Psi' - (\Psi^\dagger)'\Psi = 0$, or
$$
\vert \Psi - iL_0 \Psi' \vert^2 
- \vert \Psi + iL_0 \Psi' \vert^2 = 0\ ,
\eqn\sqrel
$$
for any $L_0 \in \R\!\setminus\!\{0\}$. Then from this one 
deduces\note{One might think that $L_0$ should be
regarded as a free parameter as in (\bcwall), but it can be
shown [\FT] that this freedom can be absorbed by adjusting 
the $U(2)$ parameters in $U$.}  
that 
$\Psi - iL_0 \Psi' = U(\Psi + iL_0 \Psi')$ with some
unitary matrix $U \in U(2)$, arriving exactly at (\safam).

In practical settings of a box, it is natural 
to impose {\it local}
probability conservation $j(0) = j(l) = 0$.
However, it is perfectly legitimate to consider
the generic boundary conditions (\safam) 
which derive from {\it global} 
probability conservation (\globalcons).  The physical
settings for those generic cases arise when,
{\it e.g.}, the two ends of the box
are combined to form a circle, creating a possible singularity
at the junction which induces a point interaction there.  
The singularity or the point interaction will then be 
characterized by the boundary conditions (\safam).
An analogous situation arises for the
system on a line $\R$ with point interaction, where
the family of self-adjoint extensions is also given by 
$U(2)$ (although the local and incomplete
parametrization $U(1) \times SL(2, \R)$ instead of $U(2)$ 
has been widely used in the literature; see 
[\ADK] for the detail).

Let us introduce a unique
parametrization for $U(2) = U(1) \times SU(2)$ by
$$
U = e^{i\xi} \pmatrix{ \alpha    & \beta \cr
                   -\beta^*  & \alpha^* \cr}\ ,
\eqn\pmt
$$
with $\xi \in [0, \pi)$ and
$\alpha$, $\beta \in \C$ satisfying
$\vert \alpha \vert^2 + \vert \beta \vert^2 = 1$, and thereby 
consider the energy spectra 
that arise under the boundary conditions (\safam).  
The general form of eigenfunctions for positive energy is
$$
\psi_k(x) = A_k\, e^{ikx} + B_k\, e^{-ikx},
\qquad k > 0\ ,
\eqn\egpos
$$ 
where the coefficients $A_k$, $B_k$ may depend on $k$.
Then, in terms of $K_\pm := 1 \pm kL_0$ 
the boundary conditions (\safam) become
$$
\pmatrix{  
\alpha K_- + (\beta e^{ikl} - e^{-i\xi}) K_+
&  
\alpha K_+ + (\beta e^{-ikl} - e^{-i\xi}) K_-
\cr
\alpha^* e^{ikl} K_+  - (\beta^* + e^{-i\xi} e^{ikl} ) K_-
&
\alpha^* e^{-ikl} K_- - (\beta^* + e^{-i\xi} e^{-ikl} ) K_+
\cr}
\pmatrix{ A_k \cr B_k \cr} = 0\ .
\eqn\bcab
$$
{}For nontrivial solutions for $A_k$, $B_k$, 
the following condition for the momentum $k$
should be fulfilled,
$$
2kL_0\,(\beta_{\rm I} + \sin\xi\, \cos{kl})
+ \left[(\cos\xi - \alpha_{\rm R}) 
+ (\cos\xi + \alpha_{\rm R})\,
(kL_0)^2 \right]\, \sin{kl} = 0\ ,
\eqn\spcpos
$$
with $\beta_{\rm I}$ and $\alpha_{\rm R}$ being
the imaginary part of $\beta$ and the real part of $\alpha$,
respectively.  
Thus the positive spectrum consists of infinitely many
discrete levels, and for large $k$
the allowed $k$ approaches to the equidistant values
$n\pi/l$ with $n$ integer.

On the other hand, for negative energy states, 
$$
\psi_\kappa(x) = A_\kappa\, e^{\kappa x} 
+ B_\kappa\, e^{-\kappa x},
\qquad \kappa > 0\ ,
\eqn\egneg
$$
one proceeds 
analogously by replacing $k$ with $-i\kappa$ in (\spcpos)
to obtain 
$$
2\kappa L_0\,(\beta_{\rm I} + \sin\xi\, \cosh{\kappa l})
+ \left[(\cos\xi - \alpha_{\rm R}) 
- (\cos\xi + \alpha_{\rm R})\,
(\kappa L_0)^2 \right]\, \sinh{\kappa l} = 0\ .
\eqn\spcneg
$$
It is readily seen by inspection that the condition (\spcneg)
has at most two
solutions for $\kappa$.
The fact that the number of
negative states is at most two 
is a norm for a system possessing
a $U(2)$ family of self-adjoint Hamiltonians,
which can be seen in the system 
$\R^1\!\setminus\!\{0\}$ as well [\CH, \ABD].
As for the existence condition for a zero energy 
state, one puts the form 
$$
\psi_0(x) = A\, x + B\
\eqn\egzero
$$
in (\safam) and finds that the condition for 
the existence reads
$$
(\beta_{\rm I} + \sin\xi) - {l\over{2 L_0}}
(\alpha_{\rm R} - \cos\xi) = 0\ .
\eqn\spczero
$$
An important point to observe 
is that the entire energy spectrum (positive, negative and
zero) depends only 
on the parameters $(\xi, \alpha_{\rm R}, \beta_{\rm I})$
rather than on the full set of $U(2)$ parameters in
(\pmt).  This suggests that, even though all the
distinct self-adjoint extensions of Hamiltonians 
are furnished by the $U(2)$ family, some of them
are unitarily related and the true space of
spectra, the spectral space ${\cal X}_{\rm SP}$, 
forms a subspace of $U(2)$.  In other words,
different self-adjoint extensions may {\it not}
correspond to different physics as opposed to
the common belief; see [\RS].
Within the $U(2)$ family there are a number of 
distinguished subfamilies (some of which are not disjoint), 
which we shall discuss 
in detail in the following.

\medskip
\noindent
{\bf Separated subfamily ${\cal F}_1$:}
The first subfamily ${\cal F}_1$ 
of our concern is one given by
the boundary conditions (\safam) with diagonal $U$.  
The boundary conditions then split into two sets, 
one at the left wall $x = 0$ and the
other at the right wall $x = l$.    
In our parametrization (\pmt), this occurs for $\beta = 0$,
leading to the subgroup $U(1) \times U(1) \subset U(2)$
given by the torus,
$$
{\cal F}_1 = S^1 \times S^1 = \left\{
(\xi, \alpha_{\rm R}, \alpha_{\rm I}, 
\beta_{\rm R}, \beta_{\rm I})\, \big\vert \, 
\xi \in [0, \pi), \, \alpha_{\rm R}^2 + 
\alpha_{\rm I}^2 = 1, \, \beta_{\rm R} = 0, \, 
\beta_{\rm I} = 0\, 
\right\}\ .
\eqn\subone
$$ 
If we write $\alpha = e^{i\varphi}$ with
$\varphi \in [0, 2\pi)$ and use the scale parameters
$L_\pm := L_0 \cot{\phi_\pm}$ with 
$\phi_\pm = (\xi \pm \varphi)/2$, we find that
the boundary conditions (\safam) reduce to
$$
\psi(0) + L_+ \, \psi'(0) = 0\ , \qquad 
\psi(l) + L_- \, \psi'(l) = 0\ .
\eqn\bcwell
$$
Clearly, this subfamily 
arises when we require local
probability conservation $j(0) = j(l) = 0$, 
and therefore it describes a box in its true sense of
the word --- the left and
the right walls of the box are disconnected physically.  
The corresponding subfamily appearing in the aforementioned
context of a line with a point interaction  
is referred to as `separated' [\ADK], from which we adopt
the name of the subfamily here.

Among this separated subfamily ${\cal F}_1$ 
are four special cases 
$(L_+, L_-) = (0, 0)$, $(\infty, \infty)$, $(0, \infty)$,
$(\infty, 0)$ in which the theory becomes free from
scale parameters (apart from $l$) 
and, consequently, solvable explicitly.
Indeed, for the first two cases 
the eigenfunctions 
$\psi_n^{(L_+, L_-)}(x)$ for 
non-negative energies 
(no negative energy state is allowed)
$$
E_n = {{\hbar^2}\over{2ml^2}} (n\pi)^2\ ,
\eqn\spsep
$$ 
are given, respectively, by  
$$
\psi_n^{(0, 0)}(x)
= \sqrt{{2\over l}}\sin{{{n \pi}\over l} x},
\qquad \hbox{for} \quad n = 1, 2, 3, \ldots,
\eqn\egfsone
$$
and
$$
\psi_n^{(\infty, \infty)}(x)
= \cases{ \sqrt{{2\over l}}\cos{{{n \pi}\over l} x}, &for 
                        $\quad n = 1$, 2, $3, \ldots$ \cr
          \sqrt{{1\over l}},         &for $\quad n = 0$. \cr} 
\eqn\egfstwo
$$
As for the Feynman kernel we find 
$$
K(b, T; a, 0)
= {1 \over{2l}}
\sum_{n = -\infty}^\infty
e^{-{i\over\hbar} E_n T}
\left(e^{i{{n\pi}\over l}(b - a)} \mp
      e^{i{{n\pi}\over l}(b + a)} 
\right)\ ,
\eqn\fkwellone
$$
where the `$-$'-sign is for $(L_+, L_-) = (0, 0)$
whereas the `$+$'-sign is for $(\infty, \infty)$.
With the help of the Poisson summation formula,
$\sum_{n = -\infty}^\infty f(n) 
= \sum_{n = -\infty}^\infty 
\int_{-\infty}^{\infty} dp\, f(p)\, e^{2\pi n i}$ 
which holds for well-behaved functions $f(x)$ 
({\it i.e.}, for $f(x) \in L_1(-\infty, \infty)$ being
continuous and of bounded variation), one 
can rewrite (\fkwellone) into
$$
K(b, T; a, 0)
= \sqrt{m\over{2\pi i \hbar T}} \sum_{n = -\infty}^\infty
\left(
e^{{i\over\hbar} {m\over{2T}}\left\{(b - a) + 2nl\right\}^2}
   \mp  
e^{{i\over\hbar} {m\over{2T}}\left\{(b + a) + 2nl\right\}^2}
\right)\ .
\eqn\fkwalltwo
$$
As on the half line, 
the result (\fkwalltwo) allows for the interpretation
that the kernel is the sum of contributions from
classical paths of two distinct classes, 
one hitting $n$-times the right and left walls equally
while the other hitting $n$-times the left and $(n\pm 1)$-times 
the right, where the signature 
of $n$ is determined according to whether the particle hits
the right wall first or not.  
The phase factor, $-1$ for $(0, 0)$ and
$+1$ for $(\infty, \infty)$, attached to
the paths in the latter class suggests that
the phase shift occurs every time the
particle hits the wall, and this is the 
only nontrivial factor for the exactness of 
the WKB approximation.

One can proceed analogously for 
the third case $(L_+, L_-) = (0, \infty)$ and for the fourth
$(L_+, L_-) = (\infty, 0)$, the latter being obtained from 
the former by the parity operation $x \mapsto l-x$,
where one has the eigenfunctions,
$$
\psi_n^{(0, \infty)}(x) 
= \sqrt{{2\over l}}\sin{{{(n + {1 \over 2}) \pi}\over l} x},
\qquad \hbox{for} \quad n = 0, 1, 2, \ldots,
\eqn\egfsthree
$$
and
$$
\psi_n^{(\infty, 0)}(x) 
= \sqrt{{2\over l}}\cos{{{(n + {1 \over 2}) \pi}\over l} x},
\qquad \hbox{for} \quad n = 0, 1, 2, \ldots,
\eqn\egfsfour
$$
of energy 
$$
E_n = {{\hbar^2}\over{2ml^2}} 
\left[\left(n + {1 \over 2}\right)\pi\right]^2\ .
\eqn\ensep
$$
(Again, no negative energy state is allowed.)
Then the Feynman kernel turns out to be 
$$
K(b, T; a, 0)
= \sqrt{m\over{2\pi i \hbar T}} \sum_{n = -\infty}^\infty
(-1)^n \left(
e^{{i\over\hbar} {m\over{2T}}\left\{(b - a) + 2nl\right\}^2}
\mp 
e^{{i\over\hbar} {m\over{2T}}\left\{(b + a) + 2nl\right\}^2}
\right)\ ,
\eqn\fkwallthree
$$
where the `$-$'-sign is for $(L_+, L_-) = (0, \infty)$
and the `$+$'-sign is for $(\infty, 0)$.
The status of the WKB exactness remains similar, where now
one observes that the phase shift occurs only when the particle
hits the left wall for $(0, \infty)$ or the right wall for
$(\infty, 0)$.
(The spectrum decomposition of the Feynman kernel 
has been obtained in [\CFG] for 
generic boundary conditions, and 
the WKB exactness for
the case $(L_+, L_-) = (0, 0)$ has been mentioned earlier
[\Schulman, \Kleinert] based on the ideas developed
in [\Buslaev, \KM].)
In passing we note that all of the Feynman kernels
obtained in closed form here can be
cast into expressions in the theta-function, 
$\vartheta_3(z, \tau) = \sum_{n = -\infty}^\infty
e^{i\pi\tau n^2 + i 2n\pi z}$.

\medskip
\noindent
{\bf Scale independent subfamily ${\cal F}_2$:}
The second subfamily ${\cal F}_2$ 
arises when the derivative terms
become decoupled from non-derivative terms in (\safam).
This occurs if and only if
$$
\det (U - I) = \det (U + I) = 0\ .
\eqn\decouple
$$
The conditions in (\decouple) state that 
the two eigenvalues for 
the matrix $U$ are $\pm 1$, and hence
$U$ is symmetric $U^\dagger = U$, as can also
be confirmed explicitly using (\pmt).
Then, from the unitarity of $U$ one deduces that
each term in (\safam) must vanish separately,
$$
(U - I)\Psi = 0\ , \qquad (U + I) \Psi' = 0\ . 
\eqn\twobc
$$
The set of $U$ satisfying (\twobc) forms 
a $U(2)/(U(1) \times U(1)) \simeq S^2$ subspace
in $U(2)$, where the second $U(1)$ is the Cartan subgroup of 
the $SU(2) \subset U(2)$.  Explicitly, the sphere is 
given by the constraint in the parameter space of $U(2)$,
$$
{\cal F}_2 = S^2 = \left\{
(\xi, \alpha_{\rm R}, \alpha_{\rm I}, 
\beta_{\rm R}, \beta_{\rm I})\, 
\big\vert \, \xi = {\pi\over 2},\,
\alpha_{\rm R} = 0, \,
\alpha_{\rm I}^2 + \beta_{\rm R}^2 + \beta_{\rm I}^2 = 1\, 
\right\}\ .
\eqn\subtwo
$$ 
This subfamily is distinguished in that the
dimensionful parameter $L_0$ drops out from the boundary
conditions, leaving the width $l$ of the well as the
only independent scale parameter.  One may therefore expect
that the theory becomes scale 
invariant in the limit $l \rightarrow \infty$ where
no scale parameter survives.

The energy spectra in this subfamily are 
characteristic,
due to the absence of the scale parameter $L_0$.  First, 
the condition (\spcneg)
does not admit any negative states, while
(\spczero)
allows one zero energy state $\psi_0(x) = \sqrt{1/l}$ 
if and only if $\beta_{\rm I} = -1$ 
(in which case $\alpha = 0$, $\beta = - i$).  Further, 
the positive energy spectra can be 
determined explicitly from (\spcpos), which reduces to
$\beta_{\rm I} + \cos{kl} = 0$, yielding the two sets of 
solutions for the momenta $k = k_s$ labeled by $s = \pm$, 
$$
k^s_n = s\, k_n\ , \qquad 
k_n := {1\over l}(\theta + 2 n \pi)\ , \qquad
n = \cases{ 0, 1, 2, \ldots, &for $s = +$, \cr
            -1, -2, -3, \ldots, &for $s = -$, \cr}
\eqn\moment
$$
where we have used 
$$
\theta := \arccos{(-\beta_{\rm I}})\ .
\eqn\agl
$$
Thus the positive energy spectra are found to be
$$
E_n = {{\hbar^2}\over{2m l^2}}(\theta + 2 n \pi)^2\ ,
\qquad
n \in \Z.
\eqn\seppos
$$
It is also readily seen that normalized eigenfunctions 
are given by 
$$
\psi_n^s(x) = A_s\, e^{ik_n^s x} - A_{-s}\, e^{-ik_n^s x}\ ,
\eqn\normeg
$$
with 
$$
A_\pm = {{(1 + \alpha_{\rm I}) 
   + (\beta_{\rm I} - i\beta_{\rm R}) e^{\mp i\theta}}\over
{2 \sqrt{l(1 + \alpha_{\rm I})
(1 +  \beta_{\rm I} \cos\theta)}}}\ ,
\eqn\coeff
$$ 
for $\alpha_{\rm I} \ne -1$ and $\beta_{\rm I} \ne \pm 1$.  
The cases $\alpha_{\rm I} = \pm 1$ in ${\cal F}_2$
are identical to the special cases $(L_+, L_-) = (0, \infty)$,
$(\infty, 0)$ of the subfamily ${\cal F}_1$, which are the
only two points of the intersection 
${\cal F}_1 \cap {\cal F}_2$.  We note that, for  
$\beta_{\rm I} = \pm 1$, the coefficients $A_\pm$ are
undetermined, which implies that all the energy levels
but one, {\it i.e.}, the zero energy 
ground state for $\beta_{\rm I} = -1$,
are doubly degenerated.  This is also seen from (\seppos)
where the two sets of levels $s = \pm$ coincide for
$\theta = 0$ and $\pi$.  
Note also that the spectra are solely
determined by the angle $\theta$, that is, 
the subfamily ${\cal F}_2$ has $S^1$ as its image
in the spectral space ${\cal X}_{\rm SP}$.
In contrast, the coefficients, and hence the eigenfunctions
are dependent on the parameters $(\alpha_{\rm I}, \beta_{\rm R})$
as well, even though the spectra are independent of them.

Having obtained the eigenfunctions in closed form, we shall
now resort the same procedure used before 
to evaluate the Feynman kernel.
To this end, let us denote by $\mathop{{\sum}'}_n$ 
the summation
$\sum_{n = 0}^\infty$ for $s = +$ and  
$\sum_{n = -1}^{-\infty}$ for $s = -$.  We then have
$$
\eqalign{
&K(b, T; a, 0) 
= \sum_{s = \pm} \mathop{{\sum}'}_n
   e^{-{i\over\hbar}E_n T} \psi_n^s(b)\, (\psi_n^s(a))^* \cr
&\quad = \mathop{{\sum}'}_n
   e^{-{i\over\hbar}E_n T}
\biggl\{
  \vert A_+ \vert^2 
  \left(e^{ik_n^+(b - a)} + e^{-ik_n^-(b - a)}\right)
+ \vert A_- \vert^2 
  \left(e^{ik_n^-(b - a)} + e^{-ik_n^+(b - a)}\right) \cr
&\qquad \qquad 
      - A_+ A_-^*
  \left(e^{ik_n^+(b + a)} + e^{-ik_n^-(b + a)}\right)
      - A_- A_+^*
  \left(e^{ik_n^-(b + a)} + e^{-ik_n^+(b + a)}\right)
\biggr\}\ .
}
\eqn\fkone
$$
Recombining the terms in the summation, 
and using the Poisson summation formula, 
we end up with
$$
\eqalign{
K(b, T; a, 0) 
&= \sum_{n = -\infty}^{\infty}
  e^{-{i\over\hbar}E_n T}  
\left(
\vert A_+ \vert^2 e^{ik_n (b - a)} 
+ \vert A_- \vert^2 e^{-ik_n (b - a)} \right. \cr
& \qquad \qquad \qquad \; \; \left.
- A_+ A_-^* e^{ik_n (b + a)} 
- A_- A_+^* e^{-ik_n (b + a)}
\right) \cr
&= 
\sqrt{m\over{2\pi i \hbar T}}\, l
\sum_{n = -\infty}^{\infty}
\biggl\{
C_n
e^{{i\over\hbar} {m\over{2T}}\{(b - a) + nl\}^2}
- D_n
e^{{i\over\hbar} {m\over{2T}}\{(b + a) + nl\}^2}
\biggr\}
}
\eqn\fktwo
$$
with
$$
C_n =
\vert A_+ \vert^2 e^{-i\theta n} 
+ \vert A_- \vert^2 e^{i\theta n},
\qquad
D_n =
A_+ A_-^* e^{-i\theta n} 
+ A_- A_+^* e^{i\theta n}\ .
\eqn\cddef
$$
The result (\fktwo) suggests 
that the WKB approximation yields the exact kernel,
if the factors $C_n$, $D_n$ are
properly interpreted.

\medskip
\noindent
{\bf Smooth subfamily ${\cal F}_3$:}
Among ${\cal F}_2$ is the
$U(1)$ subfamily ${\cal F}_3 \subset {\cal F}_2$ 
obtained by 
$$
{\cal F}_3 = S^1 = \left\{
(\xi, \alpha_{\rm R}, \alpha_{\rm I}, 
\beta_{\rm R}, \beta_{\rm I})\, 
\big\vert \, \xi = {\pi\over 2},\,
\alpha_{\rm R} = 0, \, \alpha_{\rm I} = 0, 
\beta_{\rm R}^2 + \beta_{\rm I}^2 = 1\, 
\right\}\ .
\eqn\subthree
$$ 
This subfamily is distinguished in that
with the remaining parameter 
$\theta \in [0, \pi)$ in (\agl)
the boundary conditions (\safam) become
$$
\psi(0) + e^{-i\theta} \psi(l) = 0\ ,
\qquad
\psi'(0) + e^{-i\theta} \psi'(l) = 0\ ,
\eqn\bdring
$$
which 
are actually the boundary conditions familiar on a 
smooth circle ({\it i.e.}, one without 
singularity; see [\Schulman]) with $\theta$ representing
possible phase change around a $2\pi$
rotation.  Here we find $A_+ = \sqrt{1/l}$
and $A_- = 0$, and accordingly the Feynman kernel
(\fktwo) takes the well-known form on the circle,
$$
K(b, T; a, 0) 
= \sqrt{m\over{2\pi i \hbar T}}
\sum_{n = -\infty}^{\infty}
e^{-i\theta n}
e^{{i\over\hbar} {m\over{2T}}\{(b - a) + nl\}^2} \ .
\eqn\fkring
$$
It is worth pointing out that the freedom expressed by
the $U(1)$ $\theta$-parameter, which is a prototype of 
the $\theta$-parameter of the QCD vacua,
is usually ascribed to the 
ambiguity in quantization on the circle which is
topologically nontrivial (see, {\it e.g.}, 
[\Jackiw, \Isham]). 
Here it arises
as part of the $U(2)$ family of systems allowed
on a circle, where the $U(1) \subset U(2)$ subgroup
emerges upon demanding translational
invariance of the system, that is, the smoothness
given by the boundary conditions (\bdring). 
(The role of translational invariance among
the $U(2)$ family has been remarked in [\Schulman].)

\medskip
\noindent
{\bf Isospectral subfamily ${\cal F}_4$:}
The energy spectra obtained explicitly for 
the preceding subfamilies are solutions of
the spectral condition (\spcpos) (and (\spczero)).
Conversely, one may look for specific cases where 
other solutions 
can be found directly from (\spcpos).  
One then finds, for instance, that for 
$\xi = 0$ and $\beta_{\rm I} = 0$ the condition (\spcpos)
simplifies to $\sin{kl} = 0$ and admits equidistant
$k = n\pi/l$ with $n = 1$, 2, $3, \dots$ for
the solutions.  Hence, the subfamily ${\cal F}_4$
defined by 
$$
{\cal F}_4 = S^2 = \left\{
(\xi, \alpha_{\rm R}, \alpha_{\rm I}, 
\beta_{\rm R}, \beta_{\rm I})\, 
\big\vert \, \xi = 0,\,
\alpha_{\rm R}^2 + \alpha_{\rm I}^2 + \beta_{\rm R}^2 = 1, \, 
\beta_{\rm I} = 0 \,\right\}\ ,
\eqn\subfour
$$ 
possesses the unique spectrum given by (\spsep),
even though the sphere (\subfour) retains 
$\alpha_{\rm R}$ as a free parameter for 
the spectra.  In other words, the sphere 
${\cal F}_4$ maps to a point in ${\cal X}_{\rm SP}$.  
Note that ${\cal F}_4 \cap {\cal F}_1 = S^1$ while
${\cal F}_4$ is disjoint from ${\cal F}_2$.

Despite the triviality of the spectrum, 
the eigenfunctions turn out to be nontrivial in that 
the coefficients appearing in the solution (\egpos)
become dependent on the level $n$, in sharp
contrast to the coefficients in (\normeg) in the 
scale independent subfamily ${\cal F}_2$
where they become independent (except the $s = \pm$
dependence). 
Because of this complication, it does not seem to be 
possible to proceed analogously to 
obtain the Feynman kernel in a form in which 
the WKB exactness can be examined.

\medskip
\noindent
{\bf Semi-isospectral subfamily ${\cal F}_5$:}
An extension of the isospectral subfamily ${\cal F}_4$ 
is given by those cases with the property 
$\sin \xi = \pm \beta_{\rm I}$, that is,
$$
{\cal F}_5 = \left\{
(\xi, \alpha_{\rm R}, \alpha_{\rm I},
\beta_{\rm R}, \beta_{\rm I})\,
\big\vert \, \sin \xi = \pm \beta_{\rm I}, \, 
\alpha_{\rm R}^2 + \alpha_{\rm I}^2 + 
\beta_{\rm R}^2 + \beta_{\rm I}^2 = 1 \,
\right\}\ .
\eqn\subfive
$$
Generically, in ${\cal F}_5$ 
the solutions of the spectral condition
(\spcpos) consists of two infinite sequences, 
one that is equidistant and
parameter independent
--- like for the isospectral subfamily ${\cal F}_4$ 
which arises at $\beta_{\rm I} = 0$ in ${\cal F}_5$ --- 
and another that is parameter dependent and given by  
transcendental roots.  For example, on
the positive branch 
$\sin \xi = + \beta_{\rm I}$, 
the roots of 
$\cos kl = -1$, $k = n\pi/l$ 
with $n = 1$, 3, $5, \dots$, are the parameter 
independent solutions.  By inspection, 
one finds from (\spcpos) that the other 
roots are an infinite sequence of
nontrivial, transcendental solutions. 
The two sequences of roots are
alternating, one transcendental 
root between any two succeeding equidistant
roots and vice versa.  For 
the negative branch $\sin \xi = - \beta_{\rm I}$, 
the isospectral roots are now given by the solutions, 
$k = n\pi/l$ with $n = 2$, 4, $6, \dots$, of
$\cos kl = + 1$ and the transcendental roots 
are also different from the previous ones, 
but the qualitative picture of the 
spectrum remains the same.

We observe that the two branches meet each other 
in ${\cal F}_4$ at $\beta_{\rm I} = 0$, where
both sequences become isospectral and equidistant. 
The other special
points in ${\cal F}_5$ are the two points 
with $\beta_{\rm I} = \pm 1$,
where the two sequences coincide and hence
the energy levels become doubly degenerate.
These points, $\beta_{\rm I} = 0$, $\pm 1$, are
in fact exceptional in ${\cal F}_5$ in the sense that
a single $\xi$ corresponds 
to the respective $\beta_{\rm I}$.  Since 
there exist two $\xi$ for a generic
$\beta_{\rm I}$ except the above three, we 
see from (\subfive) that, 
topologically, the subfamily ${\cal F}_5$ is given by 
two $S^3$ sharing their equators ($\beta_{\rm I} = 0$)
and the North and South Poles ($\beta_{\rm I} = \pm 1$). 
We also note that 
${\cal F}_5 \cap {\cal F}_1  
= {\cal F}_4 \cap {\cal F}_1 = S^1$, and that
${\cal F}_5 \cap {\cal F}_2 = {\cal F}_5 \cap {\cal F}_3$
consists of the two special
points $\beta_{\rm I} = \pm 1$.

\bigskip

{}Finally, we stress that 
it is important to analyze
the spectral space ${\cal X}_{\rm SP}$ thoroughly 
in order to understand fully, {\it e.g.}, 
the intriguing double spiral structure
and a certain \lq fermion-boson' duality on ${\cal X}_{\rm SP}$ 
recently reported [\Cheon, \CSb].   
We suspect that behind them underlie certain symmetries, such 
as those associated 
with parity, time-reversal and scale transformations studied
in [\ADK] for $\R^1\!\setminus\!\{0\}$.  The analysis is 
also important with regard to quantum mechanical
symmetry breaking, in view of the fact 
that in two and three dimensions 
the conformal $SO(2, 1)$ symmetry is seen to be broken 
dynamically under the presence 
of the delta-function interaction [\Jackiwb].
Our investigation on these
issues, including a fuller account of the result presented
here, will be reported elsewhere [\FT].

\bigskip
\noindent
{\bf Acknowledgement:}  
The authors wish to thank T.~Cheon and P.~\v{S}eba for 
helpful discussions.  This work is supported in part 
by the Grant-in-Aid for Scientific Research (C)
under Contract No.~11640301 provided by
the Ministry of Education, Science, Sports and
Culture of Japan.


\ve
\baselineskip= 15.5pt plus 1pt minus 1pt
\parskip=5pt plus 1pt minus 1pt
\tolerance 8000
\vfill\eject

  \vfill\eject\immediate\closeout\reffile
  \centerline{{\bf References}}\bigskip\frenchspacing%
  \input refs.tmp\vfill\eject\nonfrenchspacing
\bye